\documentclass[epj]{svjour}
\usepackage{subfigure}
\usepackage[english]{babel}
\usepackage{float}
\usepackage{slashed}
\usepackage[utf8]{inputenc}
\usepackage{hyperref}
\usepackage{amsmath}
\usepackage{latexsym}
\usepackage{lineno}
\newcommand{\bea}{\begin{eqnarray}}
\newcommand{\eea}{\end{eqnarray}}
\newcommand{\non}{\nonumber}
\usepackage{graphicx}
\usepackage{cite}
\usepackage{xcolor}
\hypersetup{colorlinks=True, linkcolor=blue,citecolor=green}
\begin{document}
\title{ Enhanced Four-Body Decays of Charged Higgs Bosons into Off-Shell Pseudoscalar Higgs and $W^\pm$ Boson Pairs in a Lepton-Specific 2-Higgs Doublet Model}

\author{Stefano Moretti \inst{1,2}  \and Muyuan Song \inst{3,4}
}                     
%
%
\institute{ School of Physics and Astronomy, University of Southampton, Southampton, SO17 1BJ, United Kingdom \thanks{\email{s.moretti@soton.ac.uk}} \and   Department of Physics and Astronomy, Uppsala University, Box 516, SE-751 20 Uppsala, Sweden \thanks{\email{stefano.moretti@physics.uu.se}} \and Center for High Energy Physics,
Peking University, Beijing 100871, China \thanks{\email{muyuansong@pku.edu.cn} (corresponding author)}
\and School of Physics and State Key Laboratory of Nuclear Physics and Technology, 
Peking University, Beijing 100871, China}

\date{Received: date / Revised version: date}
%
\abstract{We study the time-honoured decay $H^\pm\to A W^\pm$ but 
for the first time, we do so for the case of both $A$ and $W^\pm$ being off-shell, therefore computing a $1\to 4$ body decay. We show that the corresponding decay rate not only extends the reach of $H^\pm$ searches to small masses of the latter but also that the results of our implementation differ significantly from the yield of the $1\to3$ body decay over the phase space region in which the latter is normally used. We show the phenomenological relevance of this implementation in the case of the so-called
lepton-specific 2-Higgs Doublet Model (2HDM) over the mass region wherein the 
aforementioned $1\to4$ body decay can dominate just beyond the top (anti)quark mass. This mass region is accessible in the lepton-specific 2HDM as the Yukawa couplings are such that limits from  $b \to s \gamma$ and $\tau \to  \mu  \nu_{\tau} \bar{\nu_\mu}$ 
observables on $M_{H^\pm}$ are rather mild. However, we emphasize that similar effects may occur in other 2HDM types, as the $W^\pm H^\mp A$ vertex is 2HDM type independent.
}
\maketitle
%

\section{Introduction}\label{sec:in}

The discovery of a neutral Higgs boson in 2012 at the Large Hadron Collider (LHC) has been a significant breakthrough, as such a state (henceforth, denoted by $h$) is a crucial component of the Standard Model (SM) of particle physics \cite{ATLAS:2012yve,CMS:2012qbp}. In fact, the excellent agreement between SM predictions and the subsequent measurements of the $h$ properties (mass, coupling, spin, CP quantum numbers) is a remarkable achievement. In the SM, this particle emerges from a single doublet structure of a complex Higgs field, in which 4 degrees of freedom give rise to the $h$ itself and the longitudinal polarisation component of the $W^\pm$ and $Z$ bosons, following spontaneous Electro-Weak Symmetry Breaking (EWSB), see \cite{Khalil:2022toi} for a recent review. 

However, the possibility of higher Higgs representations, such as more doublets or triplets,  with more scalar or pseudoscalar Higgs states than the discovered one, has not been ruled out yet. Further, notice that the latter can include in their Higgs sector states which have a non-zero Electro-Magnetic (EM) charge, which is of significant interest due to the absence of any spin-zero charged particle in the SM and of theoretical reasons to forbid its existence. Therefore, the production and decay modes resulting from electrically charged interactions involving Higgs bosons can provide a simple way to investigate if any extended structure is underlying the observed Higgs state.
 
 We are concerned here with singly charged Higgs boson state
 ($H^{\pm}$), like those belonging to a 2-Higgs Doublet Model (2HDM)
 \cite{Branco:2011iw}. Such an extended Higgs structure is of particular interest as it can be embedded in both Supersymmetry as the Minimal Supersymmetry Standard Model (MSSM) \cite{Moretti:2019ulc, Martin:1997ns, Salam:1974yz, Ferrara:1974ac}, and Compositeness, as the Composite 2HDM (C2HDM) \cite{Mrazek:2011iu, DeCurtis:2018zvh, DeCurtis:2019jwg, DeCurtis:2021uqx}, indeed,
 two viable theories of the EW scale, i.e., that remedy the
 hierarchy problem of the SM. In particular, we want to study here
 the following charged Higgs boson decay: $H^\pm\to A W^\pm$, wherein $A$ is a CP-odd (or pseudoscalar) neutral Higgs state emerging in the 2HDM alongside two more CP-even (or scalar) ones ($h$ and $H$, with $M_h < M_H$), for a total of 5 of these.
 
The first studies of $H^\pm\to AW^\pm$ in the context of Supersymmetric models were carried out in Refs.~\cite{Moretti:1994ds, Djouadi:2005gj}, where the $1\to2$ and $1\to3$ body decay channels were examined in the MSSM. On the Compositeness side, some simple results on this were presented in Ref.~\cite{DeCurtis:2018zvh}.  However, a generic treatment of a 2HDM in relation to controlling ensuing Flavour Changing Neutral Currents (FCNCs), as the one afforded in Refs.~\cite{Branco:2011iw, Gunion:1989we, Gunion:1992hs, Pich:2009sp}, wherein different Yukawa types can eventually be mapped onto the MSSM  (a type II) and C2HDM (an aligned type), or indeed other theories, is the most common approach. 

In this connection, recent reviews on $H^\pm$ phenomenology in 2HDMs (and beyond) at the Large Hadron Collider (LHC) (and elsewhere) can be found in Refs.~\cite{Akeroyd:2016ymd, Arhrib:2018ewj}. Herein, it has been made clear that, typically, $1\to2$ and $1\to3$ body decays of the
channel $H^\pm\to AW^\pm$ are used in literature, depending on whether $M_{H^\pm}\ge M_A+M_{W^\pm}$ or $M_{H^\pm}< M_A+M_{W^\pm}$, respectively. Even in the most recent 
phenomenological studies of $H^\pm\to AW^\pm$ decays, found in Refs.~\cite{Kling:2015uba, Arhrib:2016wpw, Arhrib:2020tqk},  such a decay mode was studied in these two different kinematic 
configurations,  $1\to2$ decays (in a type-II 2HDM) for the former one and $1\to3$ body decays (in a type-I 2HDM) in the latter two\footnote{Notice that, in literature, when computing the $1\to3$ body decay, one normally allows for the $W^\pm$ boson being off-shell rather than the $A$ one, as $\Gamma_{W^\pm}\gg \Gamma_A$ over the entire parameter space of popular 2HDMs. (We will come back to this point later.)}. However, it is worth mentioning that although recent investigations conducted a parameter scan of the phenomenology involving on-shell $H^\pm\to AW^\pm$ decay in both type-I and type lepton-specific 2HDM at the LHC~\cite{Sanyal:2019xcp, Bahl:2021str}, the case when one of $A$ or $W^\pm$ is off-shell and both are off-shell has not been covered.

In this paper, we revisit these approaches to the computation of
$H^\pm\to AW^\pm$ decays, by showing that the $1\to3$ one is not the correct extension to the $1\to2$ one when $M_{H^\pm}< M_A+M_{W^\pm}$, as we will show that the most appropriate approach is always to compute the $1\to4$ body decay, wherein both $A$ and $W^\pm$ can be off-shell (separately or simultaneously). This is true not only when $M_{H^\pm}<{\rm min}(M_A,M_{W^\pm})$ but also 
when ${\rm min}(M_A,M_{W^\pm})<M_{H^\pm}<M_A+M_{W^\pm}$. To illustrate the phenomenological relevance of our approach at the LHC, we 
adopt here the so-called lepton-specific 2HDM, as this scenario  can afford one with a
$H^\pm$ state, which is relatively light (i.e., with a mass comparable to the top (anti)quark mass or lighter) and so is (necessarily) the $A$ state. 

The paper is organized as follows. The lepton-specific 2HDM is presented in the next section. We then show the constraints existing from LHC analysis on the $H^\pm\to A W^\pm$ decay in section~\ref{sec: LHCside}. Our phenomenological analysis of the
$1\to 4$ process, including how it compares to the $1\to3$ and $1\to 2$ ones over the region $M_{H^\pm}>m_t$, is reported upon  
in  section~\ref{sec: analysis}, 
We finally conclude in section~\ref{sec: con}. 

\section{Charged Higgs Bosons in the Lepton-Specific 2HDM}\label{sec:2hdmM}
In a general 2HDM, the SM is supplemented with two complex SU(2)$_L$ doublets instead of a single one, and these originate two Vacuum Expectation Values (VEVs) obeying the sum rule $v_\text{SM} =\sqrt{ v^2_1 +v^2_2}=246$ GeV, as follows:
\bea
\Phi_i=
\left(
\begin{array}{c}
     \phi^+_i  \\
      v_i + \rho^{0}_i + I \eta^{0}_i
\end{array}
\right),\qquad i = 1,2.
\eea
The physical charged Higgs states are obtained from mixing the two gauge eigenstates $\phi^+_1$ and $\phi^+_2$ by a two $\times$ two rotating matrix ($\mathcal{R}$) such that
\bea
\left(
\begin{array}{c}
     G^+  \\
     H^+
\end{array}
\right) = \mathcal{R} 
\left(
\begin{array}{c}
     \phi^+_1  \\
     \phi^+_2
\end{array}
\right),\,\,  \mathcal{R} =
\left(
\begin{array}{cc}
     \cos\beta & -\sin\beta  \\
    \sin\beta & \cos\beta
\end{array}
\right),
\eea
where $H^+$ is the mass eigenstate and $G^+$ is the Goldstone mode  eaten by $W^\pm$ following EWSB. Here, $\cos\beta$ and $\sin\beta$ can be defined by correlating the two VEVs of the Higgs doublets via $\tan\beta = v_2/v_1$. After rotating the gauge eigenstates into physical states, the charged Higgs Yukawa interactions and those between gauge bosons and charged Higgs states can be expressed as\footnote{Hereafter, we use the short-hand notations $\cos X\equiv c_X$ and $\sin X\equiv s_X$.}
\bea \label{eq:YukawatypeX2hdm}
\mathcal{L}^{H^{\pm}}_{\text{Yukawa}} \supset &-&\frac{\sqrt{2}V_{ud}}{v_{\text{SM}}} \bar{u} \left(m_u \mathcal{Y}^{H^+}_u P_L + m_d \mathcal{Y}^{H^+}_d P_R\right) d H^+ \nonumber\\&-& \frac{\sqrt{2}}{v_{\text{SM}}} m_\ell \mathcal{Y}^{H^+}_\ell \bar{\nu_\ell}_L \ell_R H^+ +\, \text{h.c.},
\eea
\bea
\mathcal{L}^{H^\pm}_{\text{gauge} } \supset &-&  \frac{g}{2} \bigg[W^+_\mu (c_{\beta-\alpha} h - s_{\beta-\alpha} H  ) \partial^\mu H^-  \nonumber\\ &-& W^+_\mu H^-\partial^\mu (c_{\beta-\alpha} h - s_{\beta-\alpha} H  ) + \, \text{h.c.}  \bigg] \non\\
&+&i \frac{g}{2} \bigg[ W^+_\mu H^- \partial^\mu A - W^+_\mu \partial^\mu H^- A \bigg] \non\\ &+& \, \text{h.c.} \label{eq:gauge and Hp},
\eea
where $u\, (d,\ell)$ represents up (down, lepton) fermions. Here,  $V_{ud}$\ is the Cabibbo-Kobayashi-Maskawa (CKM) matrix while $P_L, P_R$ are projection operators for left and right-handed spinors, respectively. Furthermore, $\mathcal{Y}^{H^+}_u,  \mathcal{Y}^{H^+}_d$ and $\mathcal{Y}^{H^+}_\ell$ are the Yukawa couplings between $H^{\pm}$ and fermions. The 2HDM without 
FCNCs~\cite{Barger:1989fj} can be classified into four types when the multiple scalars do not induce any tree-level contribution due to the presence of an exact $Z_2$ symmetry: they are named type I, II, lepton-specific (sometimes also called type X or IV) and flipped (sometimes also called type Y or III)~\cite{Branco:2011iw, Aoki:2009ha}\footnote{In the aligned 2HDM, FCNCs via Higgs boson exchanges at tree-level vanish due to the assumption that one of the Yukawa matrices for each charged
fermion is proportional to the other.}. The Yukawa couplings $\mathcal{Y}_u^{H^+}$, $\mathcal{Y}_d^{H^+}$ and $\mathcal{Y}_\ell^{H^+}$ in each type of model are characterized by the parameters $\beta$, which determines the mixing between the physical states and EW ones (mentioned previously), and $\alpha$, which determines the mixing in the neutral CP-even (scalar) sector \cite{Branco:2011iw, Gunion:1989we, Gunion:1992hs}. The values of $\mathcal{Y}^{H^+}_u, \mathcal{Y}^{H^+}_d$ and $\mathcal{Y}^{H^+}_\ell$ are expressed as 
 $\cot\beta, - \cot\beta$ and $ \tan\beta$, respectively, in the lepton-specific 2HDM \cite{Gunion:1989we,Aoki:2009ha,Branco:2011iw}. The gauge boson and charged (pseudo)scalar interactions are written in terms of the functions $c_{\beta-\alpha}$ and $s_{\beta-\alpha}$ \cite{Glashow:1976nt, Paschos:1976ay}. Finally, notice that, in contrast to the Yukawa vertices between charged Higgs boson and quarks or leptons, which involves model-dependent values for $\alpha$ and $\beta$, the interaction described by the $W^{\pm}H^{\pm}A$ vertex is model-independent since it only depends on the EW gauge coupling ($g$). 

\section{Searches for Charged Higgs Bosons via 
\texorpdfstring{$H^\pm\to AW^\pm$}{HpmAW}}\label{sec: LHCside}

The search for charged Higgs particles via fermionic decay products has been ongoing for several years, encompassing both heavy ($M_{H^\pm}>m_t$) and light ($M_{H^\pm}<m_t$) states. In the search for heavier states, the ATLAS~\cite{ATLAS:2018ntn, ATLAS:2018gfm, ATLAS:2021upq} and CMS~\cite{CMS:2015lsf, CMS:2016szv, CMS:2019bfg, CMS:2019rlz, CMS:2020imj} collaborations have focused on charged Higgs particles yielding top and bottom final states, collected at collider energies ranging from 8 TeV to 13 TeV. In terms of the gauge boson scalar mixing,  ATLAS  explored $W^{\pm} Z$ final states~\cite{ATLAS:2015edr} at $\sqrt{s}= 8$ TeV, while  CMS conducted the exact search at $\sqrt{s}= 13$ TeV~\cite{CMS:2017fgp}. 
For masses below the top quark one, both hadronic and leptonic channels have been investigated by Tevatron, LEP, and LHC searches. (Some of these searches have been reviewed in Ref.~\cite{Akeroyd:2018axd}.) The D0 and CDF collaborations at Tevatron performed a search for the process $p\bar{p}\rightarrow t\bar{t}$ where one top (anti)quark decays to $H^\pm b$ at $\sqrt{s}=1.96$ TeV. The D0 collaboration searched for disappearance modes with both leptonic and hadronic final states while the CDF collaboration tested the specific appearance mode $H^\pm\rightarrow cs$ in the mass range  80--90 GeV~\cite{D0:2009oou, CDF:2009efz}. The LEP groups ALEPH, DELPHI, L3 and OPAL finalised a combined analysis at $\sqrt{s}=189-209$ GeV for ${H^\pm}$ decays via fermionic decay modes assuming BR$(H^\pm\rightarrow\text{hadrons})$ + BR$(H^\pm\rightarrow\text{leptons})=$ 1. Eventually, LHC searches extended the mass limit on $H^\pm$ to close to $m_t$ for most such modes. A study of leptonic decay modes of $H^\pm$ states, with dominant $\tau\nu$ signatures, has been performed at $\sqrt{s} = 7, 8$ and 13 TeV in both ATLAS and CMS from $pp \to t\bar{t}$ followed by $ t \to H^{\pm} b$ \cite{CMS:2012fgz, ATLAS:2012tny, ATLAS:2012nhc, ATLAS:2014otc, CMS:2015lsf, CMS:2016szv, CMS:2019bfg, ATLAS:2018gfm}. For hadronic (di-jet) modes, the CMS collaboration searched for charged Higgs bosons decaying into $cs$ and $cb$  over different mass ranges at different centre-of-mass energies. For $H^\pm \to cs$, searches were carried out at $\sqrt{s} =$ 8 TeV ($L=$ 19.7 fb$^{-1}$) for $M_{H^{\pm}}$ in the range 90 to 160 GeV~\cite{CMS:2015yvc} and at $\sqrt{s} =$ 13 TeV ($L =$ 35.9 fb$^{-1}$) for $M_{H^\pm}$ between 80 and 160 GeV~\cite{CMS:2020osd}, assuming a full $cs$ decay mode with an exclusion limit for BR($t \to H^\pm b$) ranging from 1.68\% to 0.25\%. For $H^\pm \to cb$, CMS searched the region $M_{H^{\pm}} $ from  90 to 150 GeV at $\sqrt{s} =$ 8 TeV~\cite{CMS:2018dzl}. Furthermore, the ATLAS collaboration~\cite{ATLAS:2013uxj} conducted their first $H^\pm \to cs$ search at $\sqrt{s} =$ 7 TeV in 2011 for $M_{H^{\pm}}$ up to 150 GeV and later, using data from 2015 to 2018 with an integrated luminosity of 139 fb$^{-1}$, at $\sqrt{s} =$ 13 TeV~\cite{ATLAS:2023bzb}, performed an analysis of light charged Higgs boson production from (anti)top quark decays followed by $H^\pm \to cb$ with $60 \leq M_{H^{\pm}} \leq 160$ GeV, quite recently. This search obtained a local 3$\sigma$ (estimating a global 2.5$\sigma$) excess for $M_{H^\pm} =$ 130 GeV based on neural network for $b-$jet tagging identification. In short,
despite various experiments exploring all the above search channels for charged Higgs particles, those with mixed gauge and Higgs boson final states, specifically $AW^\pm$ ones, have not received comparable attention. Therefore, we illustrate these in some detail here.

Previous $AW^\pm$  searches at electron-positron colliders (like LEP)~\cite{ALEPH:2013htx}, via $e^+e^-\to H^+H^-$, focused on the type I 2HDM with 
$M_{H^\pm}$ up to half the collider energy with $M_A$ ranging from 10 to 70 GeV. In 2019, the CMS collaboration investigated the first search for $H^\pm$ decay to $AW^\pm$ in a $pp$ collider. The search was carried out at $\sqrt{s} =$ 13 TeV with an integrated luminosity of $35.9$ fb$^{-1}$~\cite{CMS:2019idx}. Upper limits on $H^\pm$ decay rates for a charged Higgs $H^{\pm}$ range from 100 to 160 GeV were determined for three lepton final states ($e\mu\mu$ or $\mu\mu\mu$) with $M_A$ ranging from 15 to 75 GeV. The product of BR($t \to H^{\pm} b$) $\times$ BR($H^{\pm} \to A W^{\pm}$) $\times$ BR($A \to \mu^+ \mu^-$) was limited to be between 1.9$\times 10^{-6}$ to 8.9$\times 10^{-6}$ at 95$\%$ Confidence Level (CL). Furthermore, the CMS collaboration recently conducted the first search for a heavy charge state decaying into a heavy neutral Higgs (i.e., $H$, with a mass $M_H$ larger than $m_t$) and a $W^\pm$ boson~\cite{CMS:2022jqc}. It took place with $\sqrt{s} =$ 13 TeV and used data from 2016-2018 with an integrated luminosity of $L = 138$ fb$^{-1}$. The upper limits for the product of $\sigma_{H^{\pm}}$ and BR($H^{\pm} \to H W^{\pm}$) $\times$ BR($H \to \tau^+ \tau^-$) were obtained from 0.085 pb to 0.019 pb for $M_{H^{\pm}}$ between 300 and 700 GeV. The only preliminary search by the ATLAS for the decay of $H^\pm$ into a $W^\pm$ gauge boson and a pseudoscalar $A$ state, with $A$ decaying into $\mu^+{\mu}^-$, was conducted at $\sqrt{s}=13$ TeV ($L=$ 139 fb$^{-1}$) in 2021~\cite{ATLAS:2021xhq}. The analysis was performed for $M_{H^\pm}$ values of 120, 140, and 160 GeV, with $M_A$ ranging from 15 to 75 GeV, and the product of BR($t\to H^\pm b$) $\times$ BR($H^\pm\to AW^\pm$) $\times$ BR($A\to\mu^+\mu^-$) was used to determine the observed exclusion limits of 0.9 to 6.9$\times 10^{-6}$ (with expected limits ranging from 1.6 to 9.9$\times 10^{-6}$).

Before closing this section, we emphasise that all our numerical results will be presented for parameter space points of the lepton-specific 2HDM that are compliant with the above limits (relative to the $H^\pm$ sector), specifically being consistent with the outputs of HiggsBounds~\cite{HB}, which also produces limits on the neutral Higgs sector (to which we have also adhered). Furthermore, the same constraints implemented in HiggsSignals \cite{HS}, applied to our $h$ state (with mass $M_h=125$ GeV), were also accounted for. Finally, we have checked that flavour limits are also correctly considered, complying with the \cite{superIso} outputs. 

\section{Phenomenology of \texorpdfstring{$H^{\pm} \to A\,W^\pm$}{HAW} Decays at the LHC}\label{sec: analysis}

\begin{figure*}[t!]
\centering
 \includegraphics[scale=.6]{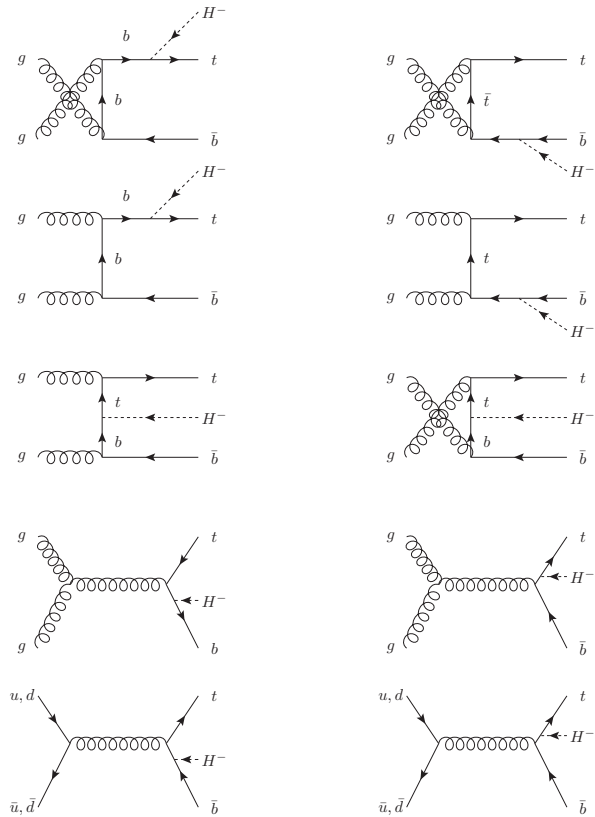}
\caption{Feynman diagrams for $gg$ (top 8 graphs) and $q\bar q$ (bottom 2 graphs) induced production of the $t\bar{b}H^-$ final state, applicable to both cases $M_{H^\pm}<m_t$ and $M_{H^\pm}>m_t$
(see Ref.~\cite{Guchait:2001pi}).}\label{fig: feynmanggtbHm}
\end{figure*}

The dominant charged Higgs production mechanism at the LHC in our scenario would mainly occur via $gg, q\bar q\to t \bar{b} H^{-}$, which Feynman diagrams are illustrated in Fig.~\ref{fig: feynmanggtbHm} (for other less significant channels for producing a charged Higgs boson at the LHC, see Ref.~\cite{Moretti:2001pp}). As explained in \cite{Guchait:2001pi}
they can be used for $H^\pm$ production, whichever mass suits our purposes as we will be scanning $M_{H^\pm}$ values around and just beyond $m_t$. Their implementation in CalcHEP~\cite{Belyaev:2012qa} has been tested (for the MSSM case) against the explicit formulae given in Refs.~\cite{Miller:1999bm, Moretti:2002eu} and deployed for the
computation in the lepton-specific 2HDM, for which we have created a dedicated model file.

Although we are focusing on the $AW^\pm$ channel in $H^\pm$ decays, the fermionic ones, mainly $\tau\nu$, $cs, cb$ and $tb$, constitute channels with significant BR, especially the first (as we are in the lepton-specific 2HDM) and last (because of the generally strong Yukawa couplings), which can be dominant over $A W^\pm$, depending on the values of $M_{H^\pm}$ and the other parameters of our scenario. One can set the heavy neutral state to be heavier than $M_{H^\pm}$, but it is crucial because the preference for a heavy scalar with $M_H \gg M_{H^\pm}$ can potentially lead to violations of electroweak precision, as defined by parameters like S, T, and U~\cite{Peskin:1991sw, Burgess:1993vc, Haber:2010bw}. Therefore, to ensure that these conditions are met, it is necessary for $M_H$ to avoid having a large mass discrepancy. One possibility to achieve this is by properly considering the $hW$ channel where the $HW$ is off, which we will explore in greater depth later on.  

\begin{figure*}[t!]
\centering
 \includegraphics[scale=1.1]{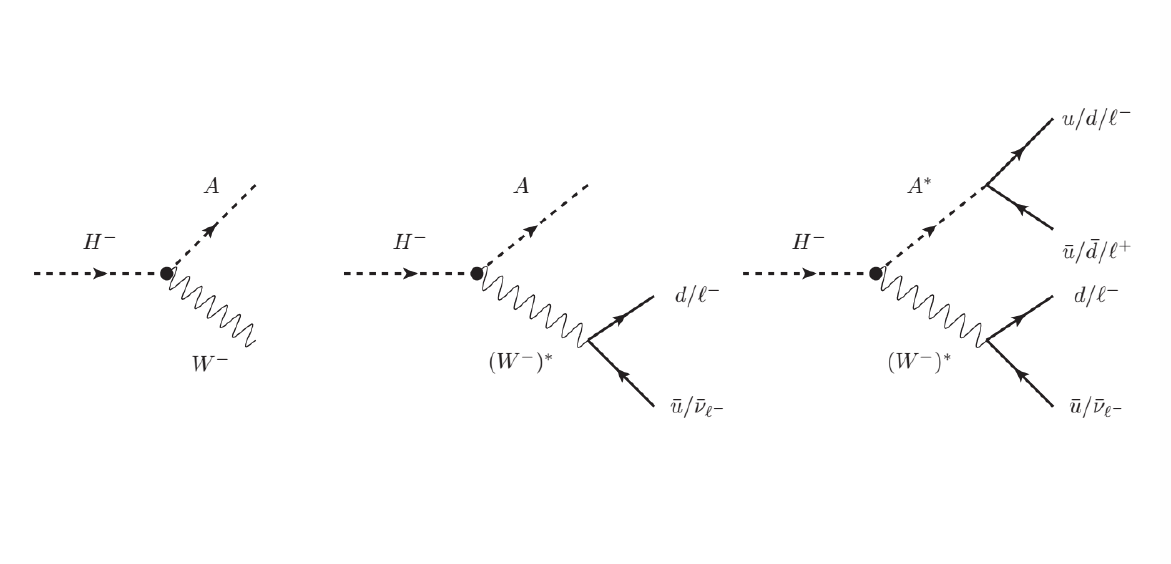}
\caption{Feynman diagrams for on shell $H^- \to A W^-$ process (left panel), three-body $H^- \to A (W^-)^*$ process (centre panel) and four body $H^- \to A^* (W^-)^*$ process with fermionic final states only (right panel). When $M_A > M_Z$, there exists a subdiagram $A \to Z \gamma$ in the top part of the right panel, which is not presented in this diagram.}\label{fig: feynmanHpAW4bd}
\end{figure*}

In the past, several studies have explored the calculations of off-shell $H^\pm \to A W^\pm$ decay, as seen in references such as Refs.~\cite{zsigmond1984three, Djouadi:1995gv, Borzumati:1998xr, Akeroyd:2002hh, Djouadi:2005gj, Moretti:1994ds}. The majority of these studies primarily focus on scenarios involving two-body and off-shell three-body decay channels, with only the last one covering the four-body decays based on the matrix elements covering the off-shell factor products~\cite{Grau:1990uu, Brown:1990th}. However, it did not delve into collider phenomenology. Thus, the derived expressions for the decay can be utilized to investigate $H^\pm \to A W^\pm$ processes at the LHC. Since the amplitude for $H^\pm AW^\pm$ provides a model-independent vertex solely on the electroweak coupling ($g$), the on-shell $H^\pm\to A W^\pm$ ($1\to2$ body) decay, depicted in the left panel of Fig.~\ref{fig: feynmanHpAW4bd} (where $M_{H^\pm}\gg M_A+M_{W^\pm}$), can be expressed as follows~\cite{Djouadi:1995gv, Djouadi:2005gj, Moretti:1994ds}:
\bea
\Gamma_{H^{\pm} \to AW^{{\pm}}}  &=& \frac{G_F}{8 \sqrt{2} \pi} \frac{M^4_{W^\pm}}{M_{H^{\pm}}} \lambda ^{\frac{1}{2}} (M^2_{A}, M^2_{W^\pm} ; M^2_{H^{\pm}}) \nonumber \\ &\times & \lambda (M^2_{A}, M^2_{ H^{\pm}} ; M^2_{W^\pm }), \nonumber \\
\lambda (x, y;z)  & =& (1 - \frac{x}{z} - \frac{y}{z} )^2 -  \frac{4xy}{z^2},
\eea
where $G_F$ is Fermi constant, $M_{H^{\pm}}, M_A$ and $M_{W^\pm}$ are the masses of the charged Higgs boson, pseudoscalar Higgs boson and $W^\pm$ boson, respectively. The aforementioned result would not hold if $M_{H^{\pm}} < M_A + M_{W^\pm}$. In this case, the ($1\to3$ body) decay $H^{\pm} \to A W^{\pm *}$ would open, wherein the gauge boson is off-shell, which subsequently decays into light fermions like in the center panel of Fig.~\ref{fig: feynmanHpAW4bd}. One method to compute the corresponding decay width is to integrate the corresponding Dalitz plot density of the process, which can be expressed as follows~\cite{Djouadi:1995gv, Djouadi:2005gj}:
\bea
&&\Gamma_{H^{\pm} \to A W^{{\pm}*} \to A f \bar{f}} = \frac{9 G^{2}_F M^4_{W^\pm} M_{H^\pm}}{16 \pi^3}\int^{1 - \frac{k_{A}}{1 - {{x}_2}}}_{1 - x_2 - k_{A}} dx_1  \nonumber\\ &&\times   \int^{1 - k_{A}}_{0} dx_2 F_{AW} , \\
&& F_{AW} = \frac{(1-x_1) (1-x_2) - k_A}{(1 - x_1 - x_2 -k_A + k_W)^2 + k_{W}{\gamma}_{W^\pm}}, \nonumber  \\
&& k_A = \frac{M^2_A}{M^2_{H^{\pm}}}, \, k_W = \frac{M^2_{W^\pm}}{M^2_{H^{\pm}}}, \, {\gamma}_W = \frac{{\Gamma}^2_{W^\pm}}{M^2_{H^{\pm}}}, \nonumber
\eea
where $M_{H^{\pm}}$, $M_A$ and $ M_{W^\pm}$ are the relevant particle masses and $\Gamma_{W^\pm}$ is the width of the $W^\pm$ boson. The integration variables ($x_1,x_2$) correspond to the energy of the final states emerging from the virtual $W^\pm$ and are expressed as $x_{1,2} = 2 E_{1,2} / M_{H^{\pm}}$. In fact, one should also account for the other $1\to 3$ body decay, i.e., $H^\pm\to A^* W^\pm$, where the Higgs boson contribution should be accounted for through the $h$ decay currents, which we have done. However, as expected, the corresponding contribution to the partial decay width is generally small, though altogether not negligible, with the $A$ width of ${\cal O}$(GeV).  

When the charged Higgs boson is lighter than both $A$ and ${W^\pm}$, then the $1\to 4$ body decay mode should be computed,
$H^\pm\to A^* W^{\pm *}$, where both $A$ and $W^\pm$ are off-shell and described by currents (as shown in the right panel of the Fig.~\ref{fig: feynmanHpAW4bd}). The ensuing partial width can be calculated through a double invariant mass squared integral~\cite{Rizzo:1980gz, Cahn:1988ru, Moretti:1994ds, Djouadi:1995gv}, 
\bea
&&\Gamma_{H^{\pm} \to A^*( W^{\pm})^*} =\int^{M^2_{H^\pm}}_{0} \frac{dq^2_A M_A{\Gamma}_A}{\pi [(q^2_A - M^2_A)^2 + ({M}_A {\Gamma}_A )^2  ]} \nonumber \\ & & \times \int^{(M_{H^\pm} - {q_A})^2}_{0}  \frac{dq^2_{W^\pm} M_{W^\pm}{\Gamma}_{W^\pm}}{\pi [(q^2_{W^\pm} - M^2_{W^\pm})^2 + ({M}_{W^\pm} {\Gamma}_W^\pm )^2  ]}  \nonumber \\ && \times    \frac{G_F}{8 \sqrt{2} \pi}  \frac{M^4_{W^\pm}}{M_{H^{\pm}}}  \sqrt{(1 - \frac{q^2_{A}}{M^2_{H^\pm}} - \frac{q^2_{W^\pm}}{M^2_{H^\pm}} )^2 -  \frac{4q^2_{A} q^2_{W^\pm} }{{M^4_{H^\pm}}} } \nonumber \\
&&  \times  \bigg [(1 - \frac{M^2_{A}}{q^2_{W^\pm}} - \frac{M^2_{H^\pm}}{q^2_{W^\pm}} )^2 -  \frac{4M^2_{A} M^2_{H^{\pm}} }{{q^4_{W^\pm}}} \bigg]\,,
\eea
where $q^2_A,q^2_{W^\pm}$ are the  (virtual) invariant masses squared of $A$ and $W^\pm$, with $\Gamma_A$ and $\Gamma_{W^\pm}$ the corresponding total widths. The two fractions within the formula, which contain $M_x\Gamma_x$ ($x=A, W^\pm$), are nothing but the well-known Breit-Wigner formulae.

All these $1\to2$, $1\to3$, and $1\to4$ decays  are calculated through a Mathematica notebook which uses a Monte Carlo (MC) routine for the numerical integrations over the phase spaces~\cite{Mathematica}. The uncertainty in the ensuing results is nominally proportional to $1/\sqrt{N}$, where $N$ is the number of evaluations performed. However, this is only true for stable
numerical integrations, so that we carefully monitored for each channel computed and decay modelling adopted that the final statistics was sufficient for our purposes, as we found the actual errors on the decay rates to be of order  $0.01\%$ in the 4- and 3-body cases (while the error is essentially zero in the 2-body one). In particular, following this, the three different implementations lead to the same result in the region $M_{H^\pm}>M_A+M_{W^\pm}$.

To obtain our final $1\to4$ body results, we had to compute the total decay width of the pseudoscalar $A$ state, for which we utilized the formulae provided in Appendix A of Ref.~\cite{Aoki:2009ha}. The total width of the $A$ state, $\Gamma_A$,  includes the tree-level decays to $q\bar{q}$ ($q=s,c$ and $b$) and $\ell\bar{\ell}$ ($\ell=\mu$ and $\tau$) as well as the one-loop decays to $gg$ and $\gamma\gamma$. For $A$ masses greater than that of the $Z$ boson, the decay $A\to Z\gamma$ must also be considered. We utilized the equations presented in Ref.~\cite{Djouadi:1995gv} to compute the on-shell decays of the heavy-charged Higgs to the top and bottom quarks and the lepton/light quark modes. However, since our interest lies in the $H^\pm$ mass range from $m_t$ to approximately 220 GeV, we also computed the $1\to 3$ body decay $H^\pm\to bt^*$, where the top (anti)quark is off-shell. To do so, we employed the formulae in eqs.~(63)-(64) of Ref.~\cite{Djouadi:1995gv}, which used the Dalitz plot density of the $H^\pm\to bt^*$ process to compute the integral with off-shell $t$ decays to $W^\pm b$. To facilitate a comparison between the analytic formula, which integrates the energy of substates (from off-shell $t$ or $\bar{t}$ ) out, and the exact integral form 
(where only $m_b$ is neglected due to its smallness compared to the energy $E_{b(\bar{b})}$), we observed a typo in the analytical expression for the $1\to 3$ body decay. Specifically, we had to omit the extra factor of 1/2 in front of Eq.~(65) to match between the analytical and the integral form. We used the HDECAY program to confirm the consistency of this off-shell decay implementation~\cite{Djouadi:1997yw, Djouadi:2018xqq}. Our findings reveal that the off-shell decay expression is about four times larger compared to the program's results. Therefore, we implemented the above formulae but with such a correction to obtain both on-shell ($1 \to 2$) and off-shell ($1 \to 3$) results for the BR($H^\pm \to t^{(*)}b$). 

As intimated, in this study, the charged Higgs boson is kept lighter than the heavy neutral scalar state, $H$, to avoid additional contributions from the decay $H^{\pm} \to H W^\pm$~\cite{Akeroyd:2022ouy, Akeroyd:1998dt}. Furthermore, in the alignment limit~\cite{Gunion:2002zf, Haber:2013mia, Asner:2013psa}, where the lightest physical Higgs state of our lepton-specific 2HDM is identical to the SM Higgs boson (i.e., $h \equiv h_{\text{SM}}$), the fact that $s_{\beta- \alpha} = 1$ would forbid the $H^{\pm} \to h W^{\pm}$ decay at tree-level. However, to ensure compliance with the electroweak oblique corrections, it becomes necessary to impose different conditions to survive the limit (i.e. The heavy scalar mass $M_H$ should not be too heavy to satisfy the S, T, U.) In this context, we disallow the $H^\pm \to H W^\pm$ channel, resulting in $s_{\beta -\alpha} = 0$, while the $H^\pm \to hW^\pm$ channel is allowed to survive the electroweak oblique observables. Consequently, the decay of the $H^\pm$ boson into a $W^\pm$ in the final state now involves both the pseudoscalar Higgs boson $A$ and the light neutral Higgs boson $h$. This approach simplifies the analysis by disregarding the $\gamma W^\pm$ and $ZW^\pm$ decays, as they are subject to one-loop suppression.

\begin{figure*}[t!]
    \centering
           \includegraphics[scale=0.52]{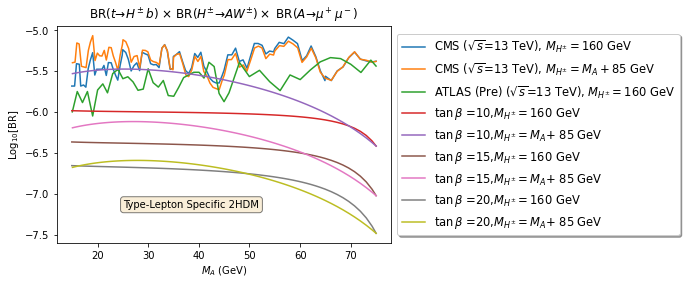}
    \caption{The exclusion bounds for the BR product in Eq.~(\ref{eq: CMSlimitproduct}) from CMS data with $\sqrt{s} =$ 13 TeV and $L=35.9$ fb$^{-1}$~\cite{CMS:2019idx} in the range of $M_A$ between 15 and 75 GeV. For the blue line we have $M_{H^{\pm}} = 160$ GeV while for the orange line we have  $M_{H^{\pm}} = M_A + 85$ GeV. The green solid line represents the upper bound from a preliminary ATLAS search (using $\sqrt{s}=13$ TeV and $L=139$ fb$^{-1}$) with $M_{H^{\pm}} = 160$ GeV~\cite{ATLAS:2021xhq}. Three BPs  ($\tan \beta = 10,15,20$) in the lepton-specific 2HDM are plotted with respective colours as in the legend, and the BR results (Y-axis) are shown in the logarithmic scale with base 10. }\label{fig: LHClimit}
\end{figure*}

 The current experimental limits from the $H^\pm\to AW^{\pm}$ search are given on the following product, 
 \bea\label{eq: CMSlimitproduct}
&& \text{BR}(t \to H^{\pm} b) \times \, \text{BR}( H^{\pm} \to A W^{\pm})\nonumber\\ & \times& \,\text{BR}(A \to \mu^+\mu^-)\,,
 \eea
where di-muon decays of the $A$ state are pursued. This is a rather sensible signature to adopt in the lepton-specific 2HDM, wherein leptonic decays of the Higgs state are generically enhanced (with respect to other 2HDM types), further recalling the efficient identification of muons and the relative cleanliness of signatures containing them. We use two experimental analyses here,
 a published CMS one ($\sqrt{s} = $13 TeV and $L= 35.9$ fb$^{-1}$~\cite{CMS:2019idx}) and a preliminary ATLAS one  $(\sqrt{s} =$ 13 TeV and $L= 139$ fb$^{-1}$~\cite{ATLAS:2021xhq}). The corresponding limits on the above product of BRs are shown in Fig.~\ref{fig: LHClimit}. In order to compare the viability of the
 lepton-specific 2HDM against such data, we have chosen here
 three Benchmark Points (BPs), with $\tan\beta = 10, 15$ and 20. 
 For the same choice of charged Higgs masses adopted by the experimental collaborations, i.e.,  $M_{H^{\pm}} = 160$ GeV (blue line) and $M_{H^{\pm}} = M_A + 85$ GeV (orange line),  we see that they are indeed compliant with data. We further note that when $\tan\beta$ is small ($\sim 8$ or less), the BR product in Eq.~(\ref{eq: CMSlimitproduct}) will be excluded for the lepton-specific 2HDM due to the large values of the BR($t \to H^{\pm} b$). This is because the Yukawa couplings between the top and bottom (anti)quarks with the charged Higgs boson are inversely proportional to $\tan\beta$, leading to the general result of smaller values of $\tan\beta$ being more strongly constrained by current bounds. On the other hand, although the branching ratio of $A \to \mu^+\mu^-$ increases with large $\tan\beta$, it remains unchanged and is subdominant compared to $A \to \tau^+\tau^-$ due to the larger mass of tau particles. This allows the BR product Eq.~(\ref{eq: CMSlimitproduct}) to survive the limits set by both CMS and ATLAS and the behavior can be observed by examining the decay width of $A$ in Fig.~2 of reference~\cite{Aoki:2009ha}. Furthermore, the low energy observables associated with $b \to s \gamma$ and $\tau \to \mu \bar{\nu} \nu$ transitions can also limit values of $M_{H^\pm}$ and $\tan\beta$ in the lepton-specific 2HDM. Concerning the former, unlike the case of type II and flipped 2HDM, where $M_{H^{\pm}}$ is required to be above 600 GeV or so,
 in the lepton-specific case only the following milder constraint applies: $M_{H^{\pm}} > 100$ GeV with $\tan\beta > 5$~\cite{Aoki:2009ha}. Concerning the latter, the decay $\tau \to \mu \bar{\nu} \nu$  imposes constraints on $\tan\beta$ such that, in our scenario, values of it greater than 20 and 40 would require $H^\pm$ to have a mass greater than 80 and 120 GeV, respectively. Thus, the three BPs chosen for our study evade these constraints too. Moreover, it is crucial to evaluate the viability of the mentioned BPs in the context of perturbative and electroweak observables. In this assessment, the HDECAY program plays a pivotal role by examining the constraints associated with these observables.
 
\begin{figure*}[t!]
    \centering
     \begin{subfigure}
    {
        \includegraphics[scale=0.48]{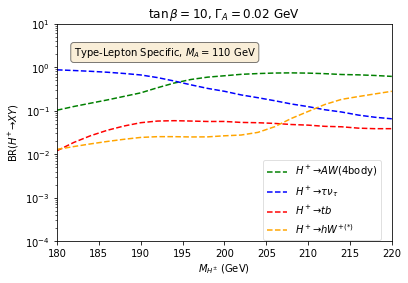}\hspace{0.4cm}
         \includegraphics[scale=0.48]{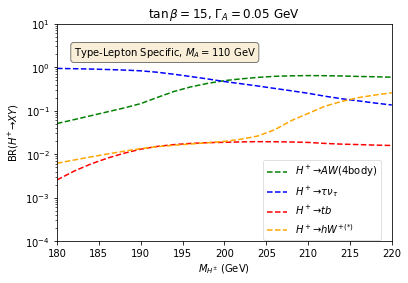}
         }
         \subfigure
        {
           \includegraphics[scale=0.48]{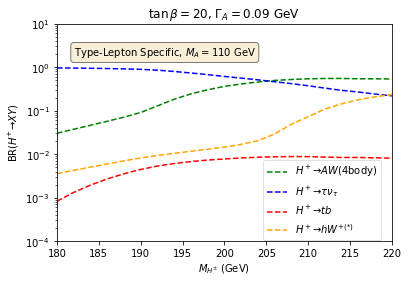}
    }
    \caption{ BR($H^{\pm} \to XY$) as a function of $M_{H^{\pm}}$ with $M_A = 110$ GeV and $\tan\beta = 10, 15$ and 20 
    (top-left, top-right and bottom frame, respectively). Here, $X$ and $Y$ correspond to three different decays of the $H^\pm$ state. In each subplot, the blue dashed line represents the BR($H^{\pm} \to \tau\nu_{\tau}$), the green dashed line relates to the BR($H^{\pm} \to A^*W^{\pm *}$), the red dashed line corresponds to the BR($H^{\pm} \to tb$) while the yellow dashed curve corresponds to the BR($H^{\pm} \to h W^{\pm*} $).
    Other decay channels ($cs, cb, ...$) have BRs below $\mathcal{O}(10^{-4})$ (i.e., they are not relevant phenomenologically) and thus are not presented here. 
    }\label{fig: BRcharHtb101520}
    \end{subfigure}	
\end{figure*}

In Fig.~\ref{fig: BRcharHtb101520}, the phenomenologically relevant decays of charged Higgs bosons (e.g., with BRs larger than $\mathcal{O} (10^{-4})$) in the lepton-specific 2HDM are presented, in the mass region 180 GeV $< M_{H^\pm} <$ 220 GeV for $M_A=110$ GeV. 
Herein, the $1\to4$ body decay is used to estimate the BR$(H^\pm\to A^*W^{\pm *})$. From these results, it is clear that  $\tau\nu_\tau$ decays generally dominate over $tb$ ones, and this pattern is obviously because our BPs have rather large values of $\tan\beta$, which then enhance the former and deplete the latter. However, it is remarkable to notice that $A^*W^{\pm *}$ decays can be very large, the more so, the bigger $M_{H^\pm}$ and the smaller $\tan\beta$ so that at times they can dominate the $H^\pm$ decay phenomenology. Another interesting spot shows the $H^\pm \to h W$ channel would be dominant than $tb$ since the $\Gamma(H^\pm \to hW^\pm)$ are forced to be constant with $\cos(\beta -\alpha)=1$ while enlarge the $\tan\beta$ would suppress the former decay\footnote{We denote the $\pm$ sign as $+$ in the figure.}. It is therefore important to see whether these parameter space regions of the lepton-specific 2HDM can become observable through this channel in the near future and, crucially, whether correspondingly the $1\to4$ body formulation of our target decay differs from the $1\to3$ and $1\to2$ ones. In fact, notice that the $H^\pm$ mass region explored in this figure is the one where both of the latter are normally used: the $1\to 2$ body decay for
$M_{H^\pm}> M_{A}+M_{W^\pm}\approx190$ GeV and the $1\to 3$ one otherwise. 

\begin{figure*}[t!]
    \centering
     \begin{subfigure}
    {
        \includegraphics[scale=0.48]{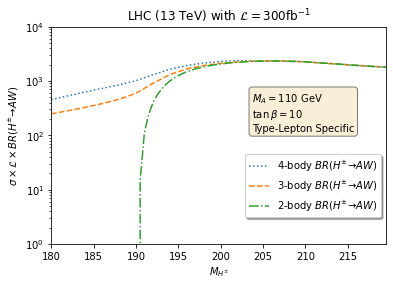}\hspace{0.4cm}
         \includegraphics[scale=0.48]{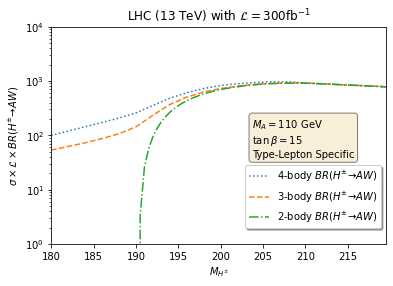}
         }
         \subfigure
        {
           \includegraphics[scale=0.48]{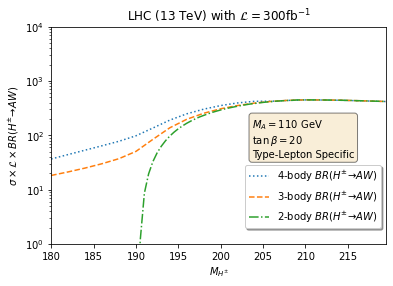}
    }
    \caption{Event rates as per Eq.~(\ref{NEvt}) at $\sqrt{s} = 13$ TeV as function of $M_{H^{\pm}}$ with $M_A = 110$ GeV and $\tan\beta = 10,15$ and 20 
    (top-left, top-right and bottom frame, respectively). Herein, the BR$(H^{\pm} \to A^{(*)}W^{\pm(*)})$ is computed as follows:
    the blue dotted line represents the $1\to4$ body case; the orange dashed line represents the $1\to3$ body case, while the green dashed-dot line represents the $1\to2$ body case. 
    }\label{fig: numevent101520300fb}
    \end{subfigure}	
\end{figure*}

With an increase in integrated luminosity up to 300 fb$^{-1}$ achievable by both ATLAS and CMS in Run 3 of the LHC \cite{Graf:2013cqj}, the number of expected $H^\pm \to A^{(*)}W^{\pm(*)}$ signal events can be expressed as a cross-section times BR times $L$ product, as follows:
\bea
&&\sigma(gg\,,q\bar{q} \to t\bar{b}H^- + {\rm c.c.}) \, \times  \, \text{BR}(H^{\pm} \to AW)\,\nonumber\\  &\times & \, \left( \frac{300~{\rm fb}^{-1}}{L}\right).
\label{NEvt}
\eea
The above expression is plotted in Fig.~\ref{fig: numevent101520300fb} as a function of $M_{H^{\pm}}$ over the usual interval,  for the $M_A$ and $\tan\beta$ choices already mentioned, 
considering $1\to2$ body (green dashed-dot line), $1\to 3$ body (orange dashed line) and $1\to 4$ body (blue dotted line) decays in the computation of BR$(H^{\pm} \to A^{(*)}W^{\pm (*)}) $. 
If $M_{H^\pm}$ is less than $M_A + M_{W^\pm}$, the $1\to2$ body decay process shuts off sharply, and no events can be generated. Here, the $1\to 3$ body and $1\to 4$ body decay results are non-zero, as expected, but the two start differing already at 220 GeV or so, with such a difference growing more and more as $M_{H^\pm}$ diminishes. Remarkably, the $1\to 4$ body results differ drastically from the $1\to3$ one below 190 GeV, well over a factor of two excess. We make this manifest in  Tab.~\ref{tab: MHpwithNum}, where the above expression is given for $M_A=110$ GeV, $\tan\beta = 10$ and in correspondence of the following choices of charged Higgs boson mass: $M_{H^{\pm}} =$ 180, 190, 200, 210 and  218 GeV.

\begin{table}[t!]
\centering
\begin{tabular}{c||c|c|c}
\hline
$M_{H^{\pm}}$ (GeV)& $1\to2$ & $1\to3$ & $1\to4$ \\
\hline
180    &0 & 248     & 450  \\
\hline
 190   &    0&    592   & 1008  \\
 \hline
 200  &    2090 & 2110   & 2277  \\
 \hline
 210 &2244 &  2245   &  2245 \\
 \hline
218 &1845 & 1846  & 1846  \\
\hline
\end{tabular}\caption{Event rates as per Eq.~(\ref{NEvt}) at
$\sqrt s=13$ TeV for sample values of $M_{H^{\pm}}$, with $M_A = 110$ GeV
and 
$\tan\beta = 10$,  depending on whether $1\to 2$, $1\to 3$ or $1\to 4$ body decays are used in the computation of BR$(H^\pm\to A^{(*)} W^{\pm (*)})$. The event numbers are rounded to the nearest integer.  
}\label{tab: MHpwithNum}
\end{table}

\section{Conclusions}\label{sec: con}
In summary, we have shown how the modelling of the decay of a charged Higgs boson $H^\pm$ into a pseudoscalar Higgs state $A$ and a $W^\pm$ gauge boson, below the threshold for on-shell $AW^\pm$ production, depends strongly upon how the off-shellness of
either or both the $A$ and $W^\pm$ states is accounted for. The naive expectation that the $1\to3$ body decay is appropriate for the description of the mass region 
min$(M_A, M_{W^\pm})<M_{H^\pm}< M_A+M_{W^\pm}$, wherein
the dominant contribution typically comes from $AW^{\pm *}$ (the
$A^*W^\pm$ channel is typically subleading as in most viable model realisations embedding such a decay one has 
$\Gamma_A\ll \Gamma_{W^\pm}$), appears to be incorrect in the presence of a complete $1\to 4$ body computation, i.e., $H^\pm\to A^* W^{\pm *}$, which yields substantially larger rates than the $1\to3$ body description in the above $H^\pm$ mass region.

These results are general, but we have illustrated them for a specific realisation of the minimal Higgs sector construct, which 
enables the aforementioned decays, i.e., a 2HDM. Specifically, we have
chosen the so-called lepton-specific realisation of it for the following reasons. Firstly, it is one for which experimental searches at the LHC have sensitivity, if anything, because these tend to exploit relatively clean signatures involving leptons (chiefly, muons) amid the overwhelming QCD background of the LHC, which are in turn generally enhanced in the 2HDM realisation chosen. Secondly, the $H^\pm\to A^{(*)} W^{\pm (*)}$ decay (when accompanied by $gg,q\bar q\to t\bar b H^-$ + c.c. production) can be phenomenologically relevant at the LHC over a substantial region of the lepton-specific 2HDM, where $\tan\beta$ is large (more than 8 or so) and $M_{H^\pm}$ is rather small (just below or above the top (anti)quark mass).   Thirdly, while such configurations of this model are currently compliant with LHC data from Run 1 and 2, which have revealed no excess in this channel, 
they could potentially become observable at Run 3, 
as event rates in the presence of a $1\to 4$ body description of the discussed decay can be up to a factor of 2 or so larger than the $1\to 3$ body ones, thereby making $H^\pm\to A^{*} W^{\pm *}$ decays a promising area of investigation for future LHC analyses.

As an outlook element, we should finish by emphasizing that we have not tested yet the discussed $1\to 4$ body decay in its natural region of validity, i.e., when $M_{H^\pm}\leq {\rm min}(M_A, M_{W^\pm})$, which we will do in a forthcoming publication, in the very low mass region of both the $H^\pm$ and $A$ states in the context of a type I 2HDM.

\section{Acknowledgements}
S.M. is  supported in part through the
NExT Institute and the STFC Consolidated Grant No. ST/L000296/1.
M.S. thanks the University of Southampton for hospitality during the initial stages of this work. We thank Andrew Akeroyd for useful comments and discussions.

\bibliographystyle{jhep}

\end{document}